\documentclass[12pt]{iopart}
\usepackage[dvipsnames]{xcolor}
\usepackage{cite}
\usepackage{dblfloatfix}
\usepackage{hyperref}
\usepackage{graphicx}
\usepackage[export]{adjustbox}
\usepackage{float}
\usepackage{xspace}

\begin{document}
\title[Selective area epitaxy of in-plane HgTe nanostructures]{Selective area epitaxy of in-plane HgTe nanostructures on CdTe(001) substrate}

\author{N Chaize$^1$, X Baudry$^1$, P-H Jouneau$^2$, E Gautier$^2$, J-L Rouvière$^2$, Y Deblock$^3$, J Xu$^{3,4}$, M Berthe$^3$, C Barbot$^3$, B Grandidier$^3$, L Desplanque$^3$, H Sellier$^5$ and P Ballet$^1$}
\address{$^1$ CEA, LETI, Univ. Grenoble Alpes, 38000 Grenoble, France}
\address{$^2$ Univ. Grenoble Alpes, CEA, INAC-MEM, 38000 Grenoble, France}
\address{$^3$ Univ. Lille, CNRS, Centrale Lille, Univ. Polytechnique Hauts-de-France, Junia-ISEN, UMR 8520 - IEMN, F-59000 Lille, France}
\address{$^4$ School of Engineering, Brown University, Providence, Rhode Island, Rhode Island 02912, USA}
\address{$^5$ Univ. Grenoble Alpes, CNRS, Institut Néel, 38000 Grenoble, France}
\ead{nicolas.chaize@cea.fr}

\begin{abstract}
Semiconductor nanowires are believed to play a crucial role for future applications in electronics, spintronics and quantum technologies. A potential candidate is HgTe but its sensitivity to nanofabrication processes restrain its development. A way to circumvent this obstacle is the selective area growth technique. Here, in-plane HgTe nanostructures are grown thanks to selective area molecular beam epitaxy on a semi-insulating CdTe substrate covered with a patterned SiO$_2$ mask. The shape of these nanostructures is defined by the in-plane orientation of the mask aperture along the $<$110$>$, $<$1$\bar{1}$0$>$, or $<$100$>$ direction, the deposited thickness, and the growth temperature. Several micron long in-plane nanowires can be achieved as well as more complex nanostructures such as networks, diamond structures or rings. A good selectivity is achieved with very little parasitic growth on the mask even for a growth temperature as low as 140\textdegree C and growth rate up to 0.5 ML/s. For $<$110$>$ oriented nanowires, the center of the nanostructure exhibits a trapezoidal shape with \{111\}B facets and two grains on the sides, while $<$1$\bar{1}$0$>$ oriented nanowires show \{111\}A facets with adatoms accumulation on the sides of the top surface. Transmission electron microscopy observations reveal a continuous epitaxial relation between the CdTe substrate and the HgTe nanowire. Measurements of the resistance with four-point scanning tunneling microscopy indicates a good electrical homogeneity along the main NW axis and a thermally activated transport. This growth method paves the way toward the fabrication of complex HgTe-based nanostructures for electronic transport measurements.
\end{abstract}
\noindent{\it Keywords\/}: Molecular beam epitaxy, Selective are epitaxy, HgTe, Nanostructures

\maketitle
\section{Introduction}

Topologically protected systems are of great interest for further developments in electronics, spintronics or quantum computing. Their ability to demonstrate exotic new physics could pave the way towards major breakthroughs for future technologies\cite{Paper1:1,Paper1:3,MolenkampScience}. 
\par
With its remarkable properties, huge \textit{g}-factor (80), low effective mass (0.02-0.03 $m_0$), high electron mobility ($>$10$^5$ cm$^2$V$^{-1}$s$^{-1}$) and strong spin orbit interaction (1.08 eV), the HgTe/CdTe system is a promising candidate to explore the potential of topological systems.  
HgTe is a II-VI semimetal with an inverted band structure that turns topological when strained to CdTe. The lattice mismatch, $f=-0.3$\%, is small enough to prevent the formation of misfit dislocations at the interface and so avoid disorder \cite{candice1,haas}, that is believed to be detrimental for quantum applications \cite{Paper17:33,Paper17:34}. The possibilities offered by HgTe for new applications and concepts have already been demonstrated in the spintronic domain \cite{MolenkampScience, SpintroniqueLeti1, SpintroniqueLeti2,HgTe} and in quantum experiments \cite{bocquillonEBL,JosephsonHgTe,HgTeEBL}.
\par 
However, one key condition to fully take advantage of these properties in nanocircuits is the one dimensional (1D) geometry including nanowires (NWs) or nanorings (NRs) \cite{fu,kitaev}. It would prevent coupling between the surface states and the bulk states of HgTe as well as limit the number of conduction channels. Therefore, the control of dimensions as well as the material quality appear to be of main importance during material growth. Vapor-liquid-solid (VLS) growth technique has originally been able to produce III-V out-of-plane nanowires with appropriate structural quality for quantum experiments \cite{KrogstrupNature,Paper2:5,Paper2:6}. However, this technique is not suitable for HgTe as the gold particles only function as nucleation sites. So the NWs suffer from poor dimension and orientation control \cite{norway,norway2}.
Moreover, this technique lacks of scalability and geometric flexibility; it requires NW manipulation for device preparation, and interconnections of only a few NWs are possible \cite{Paper6:13,Paper6:14}. Additionally, the presence of mercury makes the use of other top-down procedures difficult due to its very high volatility. Design of in-plane NWs from electron beam lithography has been proposed but this method still lacks of precise dimensional control at the nanometer scale \cite{HgTeEBL,bocquillonEBL}.
Selective area growth (SAG) has then emerged in the last years to overcome this issue and offer scalability in circuits design and integration. Here, an amorphous mask with openings designed by electron beam lithography covers a crystalline substrate. The growth only proceeds inside the openings while parasitic deposition on the mask is hindered. This approach allows the realization of complex in-plane 1D networks, with accurate control of dimensions, geometry and material quality, paving the way toward quantum networks for various materials \cite{Paper1,Paper2,Paper5,Paper6,Paper9,Paper15,Paper17,MarcusPRL}. 
Governing mechanisms of SAG for III-V materials have already been studied by several groups. The main levers are the elevation of the growth temperature and the reduction of the growth rate to avoid nucleation of adatoms on the insulating mask \cite{Paper1}. In addition, the use of an atomic hydrogen flux during the growth can enhance selectivity \cite{Paper9}. Hence, adatoms diffusion, incorporation or desorption as well as the pattern geometry (NW width, pitch or orientation) are parameters influencing the deposited thickness and so the morphology of the nanostructures \cite{Paper32,Paper33,Paper36,Paper40}. But no observation has been conducted regarding the SAG of HgTe and its driving mechanisms so far. 
\par
In this work, we present the first demonstration of selective area molecular beam epitaxy (MBE) of in-plane HgTe NW networks on semi-insulating CdTe(001) substrates. We evidence the growth of lithographically designed NW networks with well-defined cross-junctions, whose faceting depends on the opening alignment with respect to the in-plane substrate orientations. We detail the morphology of the nanostructures as function of the 2D nominal thickness, the growth temperature (GT), the openings width, W, and the pitch between them. The resulting networks are of high purity and show a sharp interface with the substrate with no sign of intermixing. A good electrical homogeneity is found along the NWs, with a thermally activated transport, signature of a bulk band gap in HgTe. HgTe on CdTe SAG shows therefore high promises for future electronic transport measurements.

\section{Experimental details}

HgTe-based nanostructures were grown in a Riber 32P chamber equipped with individual cells for Hg, Cd and Te. The Hg/Te ratio was in excess of $\sim$1000/1, following the general requirements for the growth of HgCdTe layers for infrared applications \cite{infrared}. In these conditions, the growth is only governed by the behavior of the Te atoms. The growth of the nanostructures involves several steps. First, a buffer layer of CdTe (approximately 100 nm-thick) is deposited at 300\textdegree C  under excess pressure of cadmium on the bare CdTe(001) substrate. In a different chamber, the CdTe layer is then coated with a 30 nm-thick silicon dioxyde layer that will be used as a mask. It is deposited by Plasma Enhanced Chemical Vapor Deposition at 250\textdegree C. The mask openings are then patterned with electron beam lithography according to the process described in \cite{Paper5:12}. Slits in the three major in-plane crystal directions are designed with widths ranging from 50 to 1000 nm. Crosses, rings and complex networks are also patterned. The samples are then re-introduced in the MBE chamber for the second step. After successively exposing the substrate surface to an atomic hydrogen flux and desorbing the residual surface oxide by thermal annealing up to 350\textdegree C under cadmium flux, the SAG growth of HgTe is performed. At the end of the growth process, the sample is cooled down under Hg flux to keep the surface roughness within the monolayer range, a critical condition to achieve high quality 2D layers \cite{candice1}. HgTe is grown with substrate temperature ranging from 140 to 185\textdegree C, and growth rate (GR) from 0.2 to 0.5 ML/s. 
Under these growth parameters, the desorption of Te adatoms from HgTe and CdTe crystals can be neglected. To study the influence of the GT, nanostructures are grown at 0.34 ML/s with a corresponding layer thickness of 30 nm. The NW morphology as a function of the deposited thickness is studied at GT = 170\textdegree C and GR = 0.5 ML/s. \par
After the epitaxial growth, scanning electron microscopy (SEM) and atomic force microscopy (AFM) are performed to study the morphological properties of the nanostructures. The AFM measurements were conducted in air with a Bruker Dimension ICON AFM tool. \par
The NW structural properties are studied by cross-section scanning transmission electron microscopy (STEM) in high-angle annular dark-field (HAADF) mode. The observations have been performed on a Thermo Fischer (ex FEI) Titan Themis 200 microscope equipped with spherical aberration corrector on the condenser lens for high resolution STEM imaging. A 200 kV voltage has been used for the experiments. \par 
The focused ion beam (FIB) preparation of the TEM lamellas follows a two steps procedure. First, a 50 nm-thick protective carbon layer is deposited by electron beam assisted evaporation as the direct use of ion beam would damage the nanostructure. Then, a 3 $\mu$m-thick carbon(10\%)-Pt(90\%) layer is deposited on top with the ion beam at 12 kV. The lamellas have been prepared using a Thermo Fisher Helios 5 PFIB. \par 
Transport measurements were performed with a four-probe scanning tunneling microscope (STM) under the guidance of a SEM in ultrahigh vacuum (UHV) (Nanoprobe, Omicron Nanotechnology). Prior to the transport measurements, the sample was outgassed at 80\textdegree C for a few hours and the STM tips were thoroughly annealed in the UHV preparation chamber of the Nanoprobe system. The sample consisted of NWs connected to lateral pads to ensure an easy access to the STM tips and avoid the fortuitous degradation of the narrowest NWs with the tips. The tips were first brought into electrical contact with the surface of the pads using the distance regulation of the STM control system, before being gently pushed down in the feedback-off mode. This action ensures a good and stable electric contact with HgTe, despite the thin native oxide covering the surface of the pads. Injection and collection of the current were performed through the tips connected to the most distant pads grown at the end of the NWs along their main axis. Two additional tips were connected on two other pads to measure the potential drop between both pads and then moved to the next pads. As a result, the four-probe resistance was measured for a set of different distances. The transport properties were investigated at two temperatures: 300 K (ambient) and 115 K. At the lowest temperature, the STM stage was cooled with liquid nitrogen. Because the STM tips are maintained at room temperature, the electrical contacts are usually less stable and requires a stabilization time around two hours before the measurements.

\section{Results \& Discussion}

\begin{figure}
    \centering
    \includegraphics[width = 0.9\textwidth]{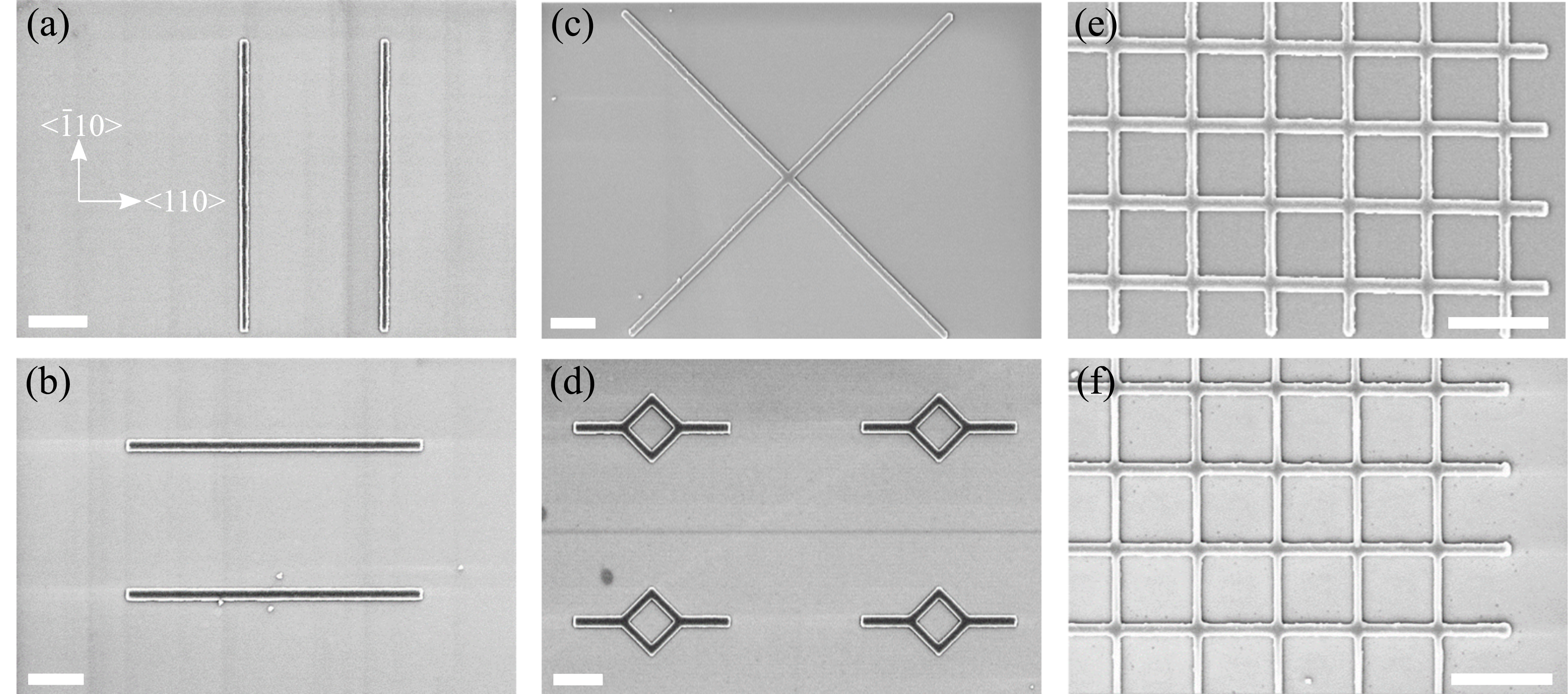}
    \caption{Top-view SEM images of HgTe nanostructures grown on a patterned CdTe(001) substrate. (a)-(b) SEM micrographs of two 5-$\mu$m-long HgTe NWs grown along, respectively, $<$$\bar{1}$10$>$ and $<$110$>$ directions. (c) SEM micrograph of a 10-$\mu$m-long $<$100$>$ type HgTe cross. (d) SEM image of diamond-like structures with junction along $<$110$>$ ridges. For images (a)-(d) the opening width is 100 nm and the growth temperature is 140\textdegree C. (e)-(f) SEM micrographs of 50 nm-wide $<$110$>$/$<$$\bar{1}$10$>$ type NW network grown at, respectively, 140\textdegree C and 180\textdegree C. The scale bar is 1 $\mu$m.}
    \label{fig:ImageMEB}
\end{figure}

\begin{figure}[ht]
    \centering
    \includegraphics[width = 0.9\textwidth]{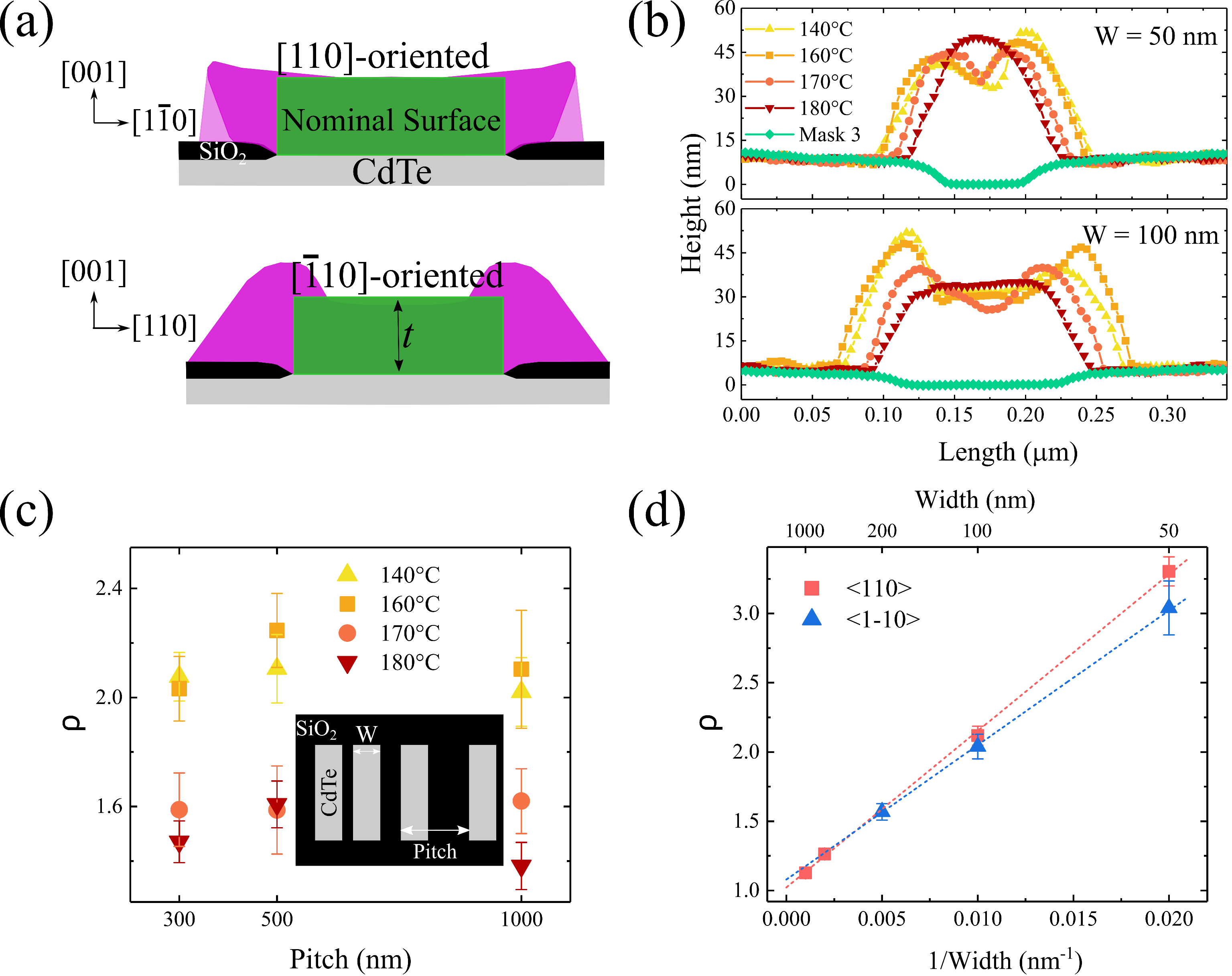}
   \caption{(a) The cross section of a nominal NW is shown in green along with typical overgrowth of experimental NWs in purple for [110] and [$\bar{1}$10] directions. The parameter $\rho$ is defined as the ratio of the whole NW cross-section (green and purple) over the green area. (b) Influence of growth temperature over the NW cross section morphology for two $<$$\bar{1}$10$>$ oriented NWs whose width is 50 nm (top) and 100 nm (bottom). The deposited thickness is 30 nm and the pitch is 1000 nm. (c) Change of $\rho$ as function of the pitch and the growth temperature for 30 nm-thick $<$$\bar{1}$10$>$ oriented NWs and width opening of W = 100 nm. (d) Evolution of experimental $\rho$ values as function of $1/\mathrm{W}$ for $<$1$\bar{1}$0$>$ and $<$110$>$ orientations types. NWs are grown at 140\textdegree C and 0.26 ML/s with a 38 nm nominal thickness and the pitch is above 1000 nm. Dashed lines are linear fit curves.}
    \label{fig:Imagerho}
\end{figure}

The SAG of HgTe nanostructures is investigated thanks to SEM observations along the high-symmetry in-plane crystal directions of the CdTe(001) substrate: $<$110$>$, $<$$\bar{1}$10$>$, and $<$100$>$.
\autoref{fig:ImageMEB}(a)-(e) display the morphologies of a 40 nm-thick growth at 140\textdegree C and $0.26$ ML/s along the three crystal directions. \autoref{fig:ImageMEB}(f), shows a 30 nm-thick growth at 180\textdegree C with a $0.34$ ML/s growth rate. The mask openings are 100 nm-wide for \autoref{fig:ImageMEB}(a)-(d) and 50 nm-wide for \autoref{fig:ImageMEB}(e)-(f). The parasitic growth of HgTe grains on the silicon dioxide mask is almost non-existent and highlights the very good selectivity of HgTe with respect to the SiO$_2$ mask during the growth even at such low growth temperatures. These are the standard temperatures normally used for the growth of 2D layers. The filling of the mask slits appears complete in every direction and the NWs have therefore a uniform shape along the entire length, up to several microns. One can see that the morphology strongly depends on the crystalline direction. While the nanostructures appear wider than the nominal opening width, revealing lateral overgrowth on the mask area, the NWs grown along $<$110$>$ direction in 
\autoref{fig:ImageMEB}(b) appear wider than the ones in \autoref{fig:ImageMEB}(a), oriented along $<$$\bar{1}$10$>$. They also seem to have more defined edges as the white rim is smoother than that observed in \autoref{fig:ImageMEB}(a). The cross along  $<$100$>$ directions, seen in \autoref{fig:ImageMEB}(c), shows good filling of the aperture and a well defined junction with no evidence of faceting between the branches. The two branches are alike and this pattern thus exhibits a fourfold-symmetric morphology. In \autoref{fig:ImageMEB}(d), the diamond shape displays good reproducibility of the complex pattern and demonstrates the good junction between $<$110$>$ and $<$100$>$ directions enabling more flexibility in nanocircuit design. Unlike the $<$100$>$ type cross from \autoref{fig:ImageMEB}(c), the networks consisting of perpendicular $<$$\bar{1}$10$>$ and $<$110$>$ directions visible in \autoref{fig:ImageMEB}(e) and (f) shows a twofold-symmetric junction due to the different growth behaviors with respect to the orientation. For 50 nm-wide NW networks in \autoref{fig:ImageMEB}(f), $<$$\bar{1}$10$>$ oriented branches appear thinner. This effect is due to an increase in the growth temperature. The underlying phenomena will be described later in \autoref{fig:Imagerho}.

\begin{figure}
    \centering
    \includegraphics[width = 0.9\textwidth]{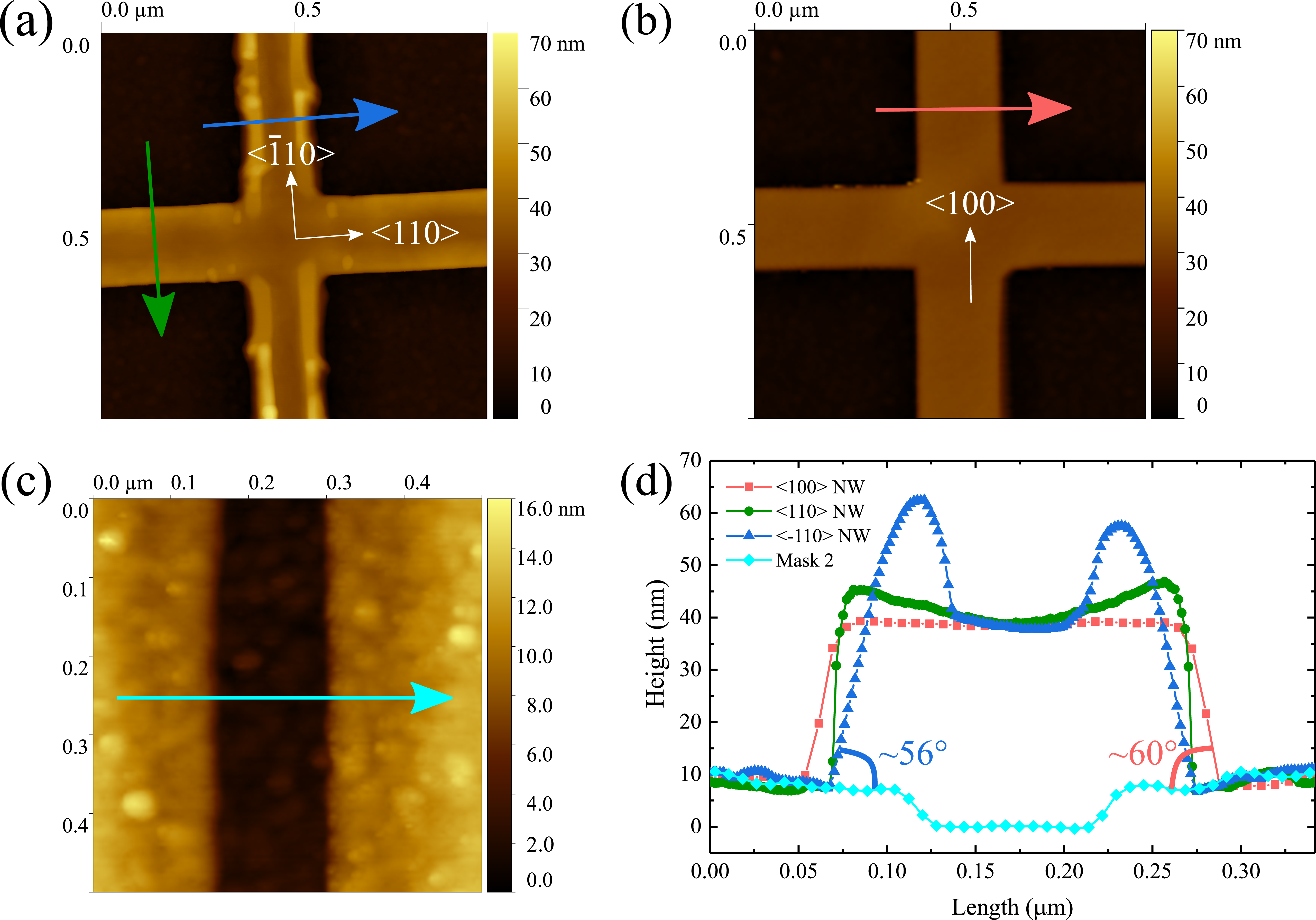}
   \caption{Nanostructures grown at 170\textdegree C and 0.26 ML/s with a 40 nm nominal thickness. AFM scans showing (a) a $<$110$>$/$<$$\bar{1}$10$>$ type and (b) a $<$100$>$ type NW cross. The darker areas in (a) and (b) correspond to the SiO$_2$ mask. Both crosses are grown from 100 nm wide slits. (c) AFM scan showing a 100 nm-wide opening before the growth. The darkest area corresponds to the CdTe substrate and the colored sides to the SiO$_2$ mask. The change in constrast across SiO$_2$ is caused by the etching process while opening the mask which reduces its thickness to about ten nanometers. (d) AFM profiles extracted from the three AFM images along the respective colored arrows.}
    \label{fig:ImageAFM}
\end{figure}

A more precise description of the nanostructures morphology is visible in \autoref{fig:ImageAFM}, where three cross sectional profiles of $<$100$>$, $<$$\bar{1}$10$>$ and $<$110$>$ oriented HgTe NWs obtained from AFM scans are displayed with respect to a mask opening profile. Nanostructures have been grown from a 100 nm-wide nominal opening at 170\textdegree C, 0.26 ML/s and a 40 nm nominal thickness. The profiles are extracted far enough from the interconnection point or from the branches tip to avoid boundary effects and so highlight the influence of the in-plane crystalline direction on the HgTe crystal growth. From the AFM profiles, a NW oriented along $<$$\bar{1}$10$>$, in blue in \autoref{fig:ImageAFM}(a) and (d), shows a $\sim$56\textdegree{} angle between the side facets and the horizontal direction, that would correspond to \{111\} facets which theoretically form a 54.7\textdegree{} angle with the (001) substrate. It is in good agreement with the study from reference \cite{Paper29}, which also showed \{111\}A facets, Hg-terminated, for this particular direction. Additionally, from the mask opening profile obtained in \autoref{fig:ImageAFM}(c), the thickness in the center corresponds to the nominal thickness of the 2D planar growth. Therefore, the two humps on the sides, forming what we name super-elevations in the following, are more likely stack of adatoms coming from the \{111\}A facets and not from the center of the NW. The origin of these super-elevations is most likely due to a kinetically driven (rather than thermodynamically) growth mode. Indeed, a longer incorporation time on the \{111\}A facet with respect to the (001) top facet would cause adatoms impinging on the \{111\}A facets to diffuse towards the  top facet. However due to the limited surface diffusion on the top facets, the diffusing atoms stay on the edges of the top surface \cite{Paper38}. In the case of a sufficiently large surface diffusion, the diffusing adatoms would cover the entire top facet, and the growth rate of the nanostructure would be higher than the nominal one as explained in \cite{Paper33}. 
In addition, a NW oriented along  $<$110$>$, in green in \autoref{fig:ImageAFM}(a) and (d), shows more vertical facets suggesting the formation of \{110\} side facets. For this direction, the super-elevations are smaller. The nominal thickness is, here as well, reached in the center. Hence, for both directions, the growth in the center seems to be alike a 2D planar growth but with particularities on the edges.
On the other side, a NW oriented along a $<$100$>$ direction, in red in \autoref{fig:ImageAFM}(b) and (d), shows strongly tilted facets oriented around 60\textdegree close to \{201\} or \{302\} side facets and a flat top surface with roughness down to 0.3-0.5 nm in good agreement with the monolayer range roughness of HgTe. The height of the NW matches the expected thickness. Comparing the last three NW profiles to the mask profile, in cyan in \autoref{fig:ImageAFM}(c) and (d), one can see that there is a significant lateral overgrowth of the nanostructures on the mask, corroborating the width of the NWs from SEM images in \autoref{fig:ImageMEB}. The lateral overgrowth and the presence of the super-elevations, where the thickness is bigger than the nominal 2D growth, suggest a diffusion of the adatoms from the mask towards the openings during the growth, as discussed later.

\begin{figure}
    \centering
    \includegraphics[width = 0.9\textwidth]{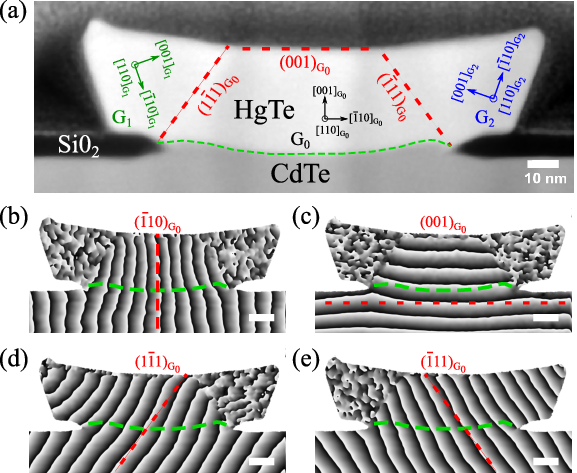}
   \caption{(a) High resolution HAADF-STEM image of the transversal cross-section of a [110]-oriented NW. We define a frame of reference for each grain with respect to its orientation. The two red 54.7\textdegree -tilted dashed lines correspond to grain boundaries of \{111\}B facets between the center grain G$_0$ and the two side grains G$_1$ and G$_2$. The top red dashed line represents the (001)$_{\mathrm{G}_0}$ plane. The dashed green curve highlights the interface between CdTe and HgTe. From this image the morphology and strain of the NW cross-section can be determined especially thanks to the numerical g-moiré images obtained by visualizing the respective (b) (1$\bar{1}$0)$_{\mathrm{G}_0}$, (c) (001)$_{\mathrm{G}_0}$, (d) (1$\bar{1}$1)$_{\mathrm{G}_0}$$\Leftrightarrow$($\bar{1}$11)$_{\mathrm{G}_1}$ and (e) (1$\bar{1}$1)$_{\mathrm{G}_0}$$\Leftrightarrow$(1$\bar{1}$1)$_{\mathrm{G}_2}$ planes. The scale bar in g-moiré maps represents 20 nm.}
    \label{fig:ImageTEM1}
\end{figure}

\begin{figure}
    \centering
    \includegraphics[width = 0.45\textwidth]{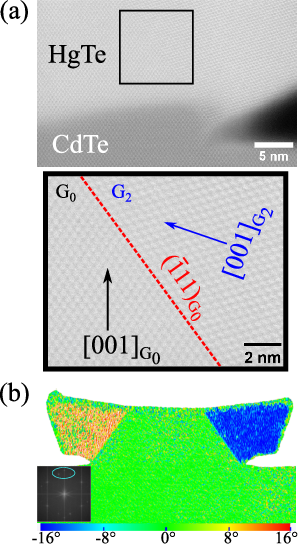}
   \caption{(a) Zoom on the crystal/mask interface near grains G$_0$ and G$_2$ from the HR HAADF-STEM image of \autoref{fig:ImageTEM1}(a). The area outlined by a black square is further zoom-in to visualize the atomic columns and clearly shows the \{111\} twin plane separating grains G$_0$ and G$_2$ that are rotated from each other by 70.5\textdegree ~(angle between [001]$_{\mathrm{G}_0}$ and [001]$_{\mathrm{G}_2}$). (b) Semi-quantitative map of the rotation angle of the filtered area in the FFT of the image (the FFT and the elliptic mask are visible in the inset). The three points with highest intensity in this area correspond to planes (1$\bar{1}$1)$_{\mathrm{G}_1}$, (001)$_{\mathrm{G}_0}$ and ($\bar{1}$11)$_{\mathrm{G}_2}$. The reference plane is taken as (001)$_{\mathrm{G}_0}$. The rotation between planes (1$\bar{1}$1)$_{\mathrm{G}_1}$, respectively ($\bar{1}$11)$_{\mathrm{G}_2}$, and (001)$_{\mathrm{G}_0}$ should  be equal to +15.8\textdegree{} (orange area), respectively -15.8\textdegree{} (blue area ).}
    \label{fig:ImageTEM2}
\end{figure}

The structural properties of a [110] oriented HgTe NW were further characterized by High Resolution (HR) STEM. The observed NW is prepared from the same sample as that presented in \autoref{fig:ImageAFM}.  The nominal slit width is 100 nm. \autoref{fig:ImageTEM1}(a) shows a high-angle annular dark-field (HAADF) STEM image of  the NW cross-section perpendicular to the [110] ridge direction. Due to the approximate Z$^2$ dependence of the HAADF image contrast, Z being the average atomic number of the local material, the “Z-lighter” silicon dioxide mask appears black with respect to the “Z-heavier” CdTe substrate and the HgTe nanostructure. The atomic columns of the image are not visible here due to reduced resolution of the figure but we could identify three grains in the NW crystal structure. We therefore define three frames of reference assigned respectively to these three grains, the G$_0$ frame corresponds to the substrate orientation. This STEM image reveals the symmetrical shape of the NW with a flat (001)$_{\mathrm{G}_0}$ top facet in the center, as suggested from the top-view SEM micrographs in \autoref{fig:ImageMEB}(b) and AFM scans in \autoref{fig:ImageAFM}(a) and (d). Two side grains G$_1$ and G$_2$ are found to extend above the mask and have a rotated surface. Thus, the edges of the NW form an overhang above the insulating mask. This geometry cannot be revealed by AFM scans, which suggested vertical facets in \autoref{fig:ImageAFM}(d) due to the tip geometry. 
Zooms on the HAADF image (see for instance \autoref{fig:ImageTEM2}(a)) clearly indicates the presence of a \{111\}B twin boundary between grains G$_0$, G$_1$ and G$_2$. The polarity of the \{111\}B facet, Te-terminated, is consistent with the work presented in reference \cite{Paper29}. The formation of these grains could be explained by the asymmetric (2$\times$1) surface reconstruction of the top (001) facet during growth \cite{Paper20, Paper21}. This asymmetric unit cell is elongated along the [1$\bar{1}$0] direction, hindering surface diffusion along this direction on the top facet. Therefore, spreading adatoms are rather forced to form a grain and grow laterally. Moreover, one can see that the substrate inside the mask opening shows two humps on the sides. The origin of their formation is not explained yet and further investigations are needed.

To provide a visual representation of these grains some particular Geometrical Phase Analysis (GPA) treatements were performed on these images. Either g-moiré images, \autoref{fig:ImageTEM1}(b)-(e), or rotation map, \autoref{fig:ImageTEM2}(b), were calculated; here g indicates the reciprocal vector that characterizes a given crystallographic plane. The principle of these treatments is to perform a Fast Fourier Transform (FFT) of the HR STEM image and apply a mask in the Fourier transform in order to select diffraction spots. For the g-moiré images, we select a region around a given g-vector associated to a given family of planes \cite{moire1, moire2}. For the rotation map, three diffracted peaks, each of them being associated to one of the three grains G$_0$, G$_1$ and G$_2$, are filtered. The size of the mask is set to preserve a balance between a good spatial resolution and a high signal-to-noise ratio. In  \autoref{fig:ImageTEM1}, four g-moiré images have been obtained with four different g-planes: (b) (001)$_{\mathrm{G}_0}$, (c) ($\bar{1}$10)$_{\mathrm{G}_0}$, (d) ($\bar{1}$11)$_{\mathrm{G}_0}$ and (e) (1$\bar{1}$1)$_{\mathrm{G}_0}$. G-moiré maps reveal the three grains and give a visual representation of the bending and interplanar distance variations of these crystallographic planes. 
Vertical (1$\bar{1}$0)$_{\mathrm{G}_0}$, \autoref{fig:ImageTEM1}(b), and horizontal (001)$_{\mathrm{G}_0}$, \autoref{fig:ImageTEM1}(c), planes are only present in the substrate and HgTe NW G$_0$ grain; so fringes are only present in G$_0$, while the grains G$_1$ and G$_2$ appear noisy. \autoref{fig:ImageTEM1}(b) shows that the vertical lattice fringes in the HgTe NW are smaller than in the substrate, indicating that the NW is partly relaxed and no longer matches the in-plane lattice parameter of the substrate. Very importantly here, there is no fringe discontinuity in the g-moiré map indicating that there is no misfit dislocation in the structure and so a good epitaxial relation exists between the CdTe substrate and the HgTe nanostructure with a continuous atomic lattice. In the scope of transport experiments, this absence of disorder and defects at the susbtrate/HgTe interface is very promising as mentioned in \cite{Paper17:33,Paper17:34}. In \autoref{fig:ImageTEM1}(c), there is not a clear difference between horizontal fringes of HgTe NW G$_0$ grain and the substrate. This is consistent with the previous g-moiré maps as a reduction of the in-plane lattice parameter of the NW yields to an increase of its out-of-plane lattice parameter, being closer to that of CdTe. In addition, inclined (1$\bar{1}$1)$_{\mathrm{G}_0}$ planes are only visible in grains G$_0$ and G$_1$ as seen in \autoref{fig:ImageTEM1}(d), and inclined ($\bar{1}$11)$_{\mathrm{G}_0}$ planes are only visible in grains G$_0$ and G$_2$ as shown in \autoref{fig:ImageTEM1}(e). Thus, each grain shares a common \{111\} plane with the central region G$_0$ and the susbtrate. The side grains therefore appear as grain G$_0$ but tilted by $\pm$70.5\textdegree, the $<$221$>$ direction in the grains is now parallel to the [001]$_{\mathrm{G}_0}$ substrate orientation. Here as well, these maps do not show any fringe discontinuity. Moreover, we notice that the \{111\} planes in the HgTe NW are slightly tilted or bent compared to those in the substrate. The relaxation would then not be equivalent between the [001] out-of-plane and [$\bar{1}$10] in-plane directions in G$_0$. Further investigations are needed to explain in detail the relaxation mechanisms inside the mask opening.

In \autoref{fig:ImageTEM2}(a), the twin plane boundary between the center (G$_0$) and the right (G$_2$) grain is shown. The common \{111\}B plane is marked with a red dashed line. The STEM image evidences the absence of stacking fault in the vicinity of the grain boundary, the substrate/nanowire interface and the nanowire/mask interface. In addition, HgTe has a crystalline structure up to the mask area and intermixing between CdTe and HgTe is limited to a few monolayers around the interface. In \autoref{fig:ImageTEM2}(b), a zone from the FFT is selected (see the blue ellipse in the inset) where the planes with highest intensity are, from left to right, ($\bar{1}$11)$_{\mathrm{G}_2}$, (001)$_{\mathrm{G}_0}$ and (1$\bar{1}$1)$_{\mathrm{G}_1}$. Then, one can access the planes rotation with respect to the horizontal direction. While the center area is found free of rotation (consistent with the moiré image of \autoref{fig:ImageTEM1}(b)), the G$_1$ grain, respectively G$_2$ grain, exhibits a +15\textdegree, respectively -15\textdegree, rotation. This is consistent with the STEM micrograph from \autoref{fig:ImageTEM2}(a), where the tilted ($\bar{1}$11) plane of the G$_2$ grain makes a $\sim$15\textdegree{} angle with the horizontal direction.

\begin{figure}[ht]
    \centering
    \includegraphics[width = 0.9\textwidth]{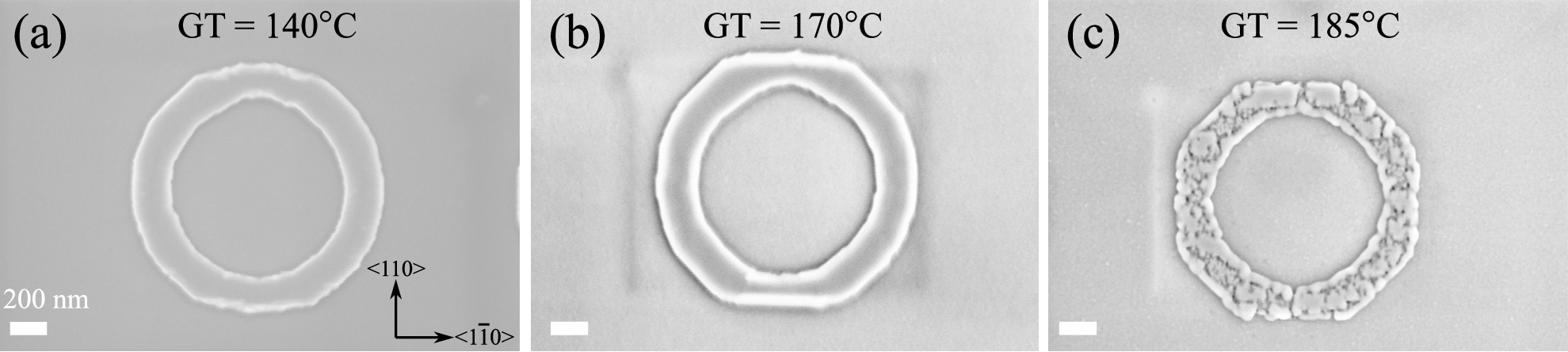}
   \caption{Top-view SEM images of three HgTe ring nanostructures grown at (a) 140\textdegree C, (b) 170\textdegree C, (c) 185\textdegree C.}
    \label{fig:ImageRing}
\end{figure}

The next parameter investigated is the growth temperature. The three top view SEM images in \autoref{fig:ImageRing} illustrate the influence of this parameter on the morphology of a 30 nm-thick ring nanostructure grown at 0.26 ML/s. The higher the GT the more visible the side faceting, the outer rim turning into an octagon. At GT = 140\textdegree C, no clear side facet is visible, only accumulation of adatoms at the edges parallel to the $<$1$\bar{1}$0$>$ direction (top and bottom part of the ring here) is observable as there is a rougher and wider outline. This effect is weaker in the inner side, at the $<$1$\bar{1}$0$>$ edges. Indeed, the collection area of diffusing adatoms for the inner rim is smaller due to its concave shape. At GT = 170\textdegree C, the \{111\} facets along the $<$1$\bar{1}$0$>$ direction are visible on the external side. Facets along the $<$110$>$ and $<$100$>$ directions, are still in the process of formation. Regarding the inner side, there is a slight adatom concentration at the $<$1$\bar{1}$0$>$ edges, similarly to the previous image. At GT = 185\textdegree C, the facets are formed in every azimuth of the plane for the outer rim. The inner side of the ring, shows no clear faceting and, therefore, keeps a more or less constant shape throughout the different growth temperatures. This is most likely because the growth is kinetically driven rather than thermodynamically, a thermodynamic growth favoring symmetric morphologies between the outer and inner rings,  as explained in \cite{Paper41}. Here, the inner rim would tend to form different side facets than the outer rim but they are likely not stable, yielding to this constant rounded shape of the inner rim for every GT. The top surface however becomes very rough. This growth temperature is then detrimental in the scope of transport measurements.

In the following part, we study the influence of the growth temperature on the shape of the NWs, mainly how the GT impacts the lateral overgrowth of the structures and their thickness.
In \autoref{fig:Imagerho}(a) two schematics present the method to compute $\rho$, the ratio between the NW cross section area (purple plus green areas) and the cross sectional area of a nominal NW ($S_{\mathrm{nom}}$ in green). We assume that the nominal NW grows layer-by-layer with vertical sidewalls up to the nominal planar growth thickness, \textit{t}. The lighter purple areas for [110] oriented ridges correspond to the additional surface calculated from AFM scans due to the overhang of the side grains. The $\rho$ variable will be used to characterized the NWs overgrowth.\\
The change in the AFM cross section profile for two $<$$\bar{1}$10$>$ oriented NW widths (W = 50 nm top and W = 100 nm bottom) with respect to the GT is displayed in \autoref{fig:Imagerho}(b). In both cases, the wider cross sections for NWs grown at 140 and 160\textdegree C are clearly visible while the height in the center remain close to the 30 nm nominal thickness. The growth performed at 180\textdegree C shows a different behavior. Indeed, as the lateral overgrowth is reduced with an increasing temperature, super-elevations disappear at 180\textdegree C. For W = 50 nm, they merge leading to a triangular shape with \{111\}A facets and a rounded tip. For W = 100 nm, a flat (001) top facet is observed. This is due to an enhanced diffusion on the crystal \cite{Paper38} yielding a higher thickness than the nominal one, consistent with the work presented in \cite{Paper33}. \\
\autoref{fig:Imagerho}(c) displays the evolution of $\rho$ as function of the pitch at four different growth temperatures. Values are calculated from $<$$\bar{1}$10$>$ oriented NWs grown from 100 nm-wide openings with a nominal 30 nm-thick deposition. The inset presents the pattern geometry designed to study the diffusion length of Te adatoms on the mask, $\lambda_{\mathrm{Te/mask}}$, Hg adatoms playing no role in the growth rate. First, one can notice that $\rho$ values are always above 1; meaning that we have more material deposition than the nominal growth and so a net diffusion from the NWs to the mask is prohibited. This regime is different from most of other works, mainly on GaAs, whose $\rho$ value is below 1; the net flux is from the growing crystal towards the mask area \cite{Paper36,Paper32}. Considering now the change in GT, we can see that NWs grown at 140 and 160\textdegree C show significantly higher $\rho$ value in comparison with growths performed at 170 and 180\textdegree C as also visible on \autoref{fig:Imagerho}(b). At GT = 180\textdegree C, the overgrowth seems to be the most limited even though there is a bit of overlapping with the NWs grown at 170\textdegree C. This change as function of the temperature can be explained by the competing phenomena at play during the growth. We assume here that there is no adatom desorption from the surface of the crystalline NW in the studied GT range, the Te sticking coefficient on HgTe or CdTe remaining 1. This assumption is corroborated by the nominal thickness found in the center of the NW for GTs below 180\textdegree C in \autoref{fig:Imagerho}(b). Hence, Te adatoms arriving onto the mask have three different behaviors: 1) nucleate on the mask, or 2) diffuse toward the HgTe crystal, or 3) undergo desorption. The change in $\rho$ therefore highlights the change in $\lambda_{\mathrm{Te/mask}}$ which is function of the diffusion coefficient $D$ and the characteristic time for desorption $\tau$. Either $\lambda_{\mathrm{Te/mask}}$ is large and more adatoms around an opening can participate in the nanostructure growth, or, as we do not observe parasitic clusters on the mask, tellurium adatoms desorb close to their impinging location due to a limited diffusion length and so less adatoms reach the opening.

\begin{figure}[ht]
    \centering
    \includegraphics[width = 0.45\textwidth]{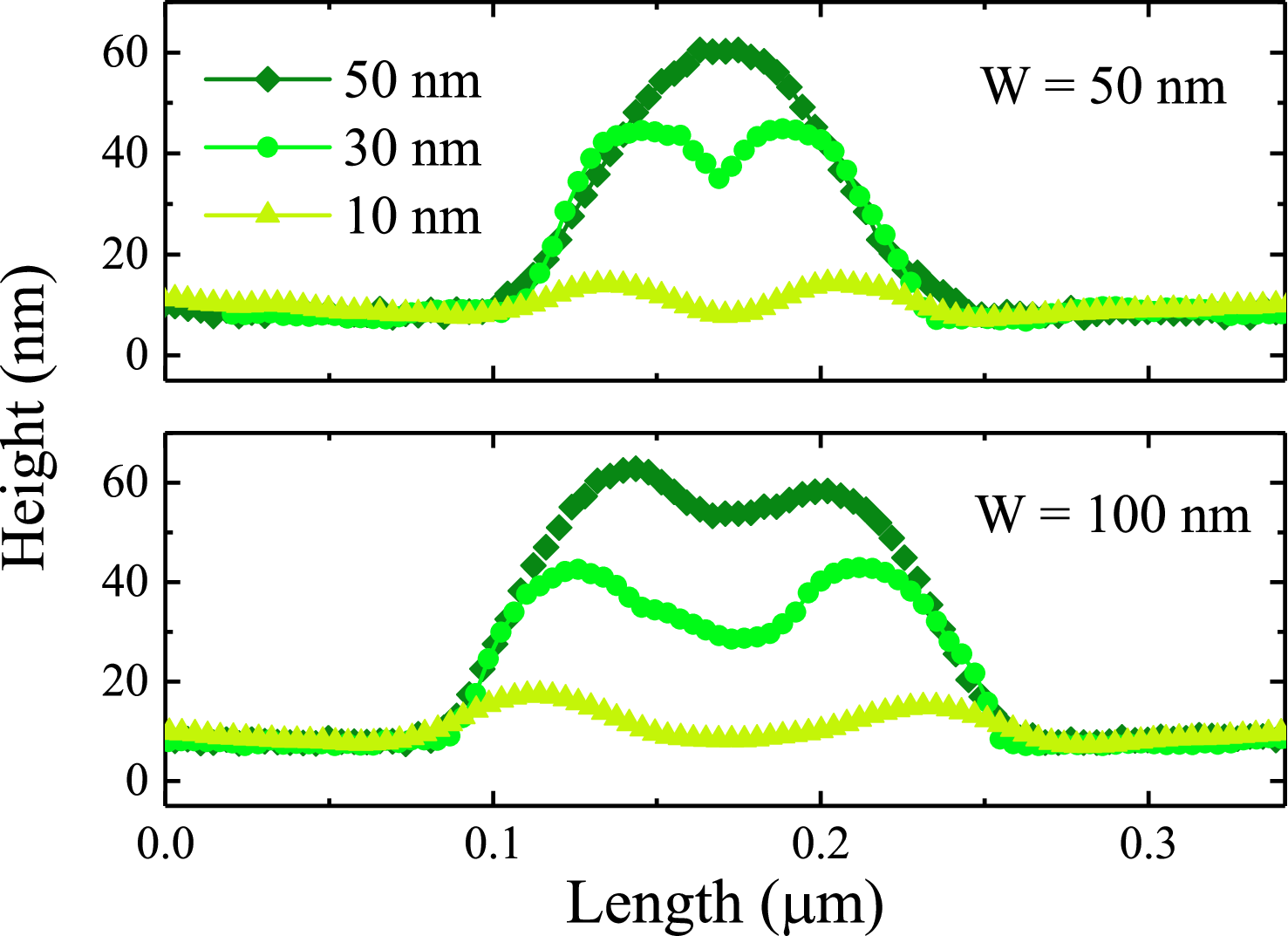}
   \caption{Change in the NW cross section morphology for three different nominal thicknesses for two $<$1$\bar{1}$0$>$ oriented NWs whose opening width is 50 nm (top) and 100 nm (bottom). The growth temperature is 160\textdegree C for the 10 nm-thick growth and 170\textdegree C for the 30 and 50 nm-thick growths.}
    \label{fig:Imagethick}
\end{figure}

In addition, as no significant change in $\rho$ with the pitch is visible (the slight fluctuations are most likely due to variations in the etching profile of the mask), $\lambda_{\mathrm{Te/mask}}$ is below $(\mathrm{\textit{min}}(\mathrm{Pitch})-W)/2$. Hence we find $\lambda_{\mathrm{Te/mask}} < 100$ nm for all studied GTs. Consequently, there is no synergetic growth effect in arrays of NWs, differing from works involving the SAG of III-V materials \cite{Paper36,Paper32}. In conclusion, the GT plays a role on the morphology of the NW by influencing the diffusion length of adatoms coming from the SiO$_2$ mask and the diffusion length of Te adatoms on the HgTe (001) facet. 

It is also interesting to characterize the overgrowth with respect to the opening width. In \autoref{fig:Imagerho}(d), the evolution of experimental $\rho$ values are plotted as function of the inverse of the NW width. If the same amount of overgrowth is assumed regardless of the opening width, one gets:
\begin{eqnarray}
    S_{\mathrm{nom}, i}(\rho_{i}-1) = S_{\mathrm{nom}, j}(\rho_{j}-1)
\end{eqnarray}
And considering that $S_{\mathrm{nom}, i}=t\cdot \mathrm{W}_i$, $\rho_{i}$ should be a linear function of $1/\mathrm{W}_i$: 
\begin{eqnarray}
    \rho_{i} = 1+( \rho_j  - 1) \frac{\mathrm{W}_j}{\mathrm{W}_i}
\end{eqnarray}  

Here we consider a set of NWs whose opening width varies from 50 to 1000 nm. \autoref{fig:Imagerho}(d) shows that the $\rho$ values computed from the experimental data of NWs in both directions increase linearly with $1/\mathrm{W}$ as there is a good alignment between the computed $\rho$ values and the fitting curve. The overgrowth is therefore found to be width independent (as long as the super-elevations do not merge), unlike most observations in other material systems \cite{Paper40, Paper32, Paper36}. The nanostructures are then relatively more selective the wider they are. One can also notice the values for $<$110$>$ oriented slits are larger. This might be due to the area below the overhang which is not subtracted in the calculation.

\begin{figure}
    \centering
    \includegraphics[width = 0.45\textwidth]{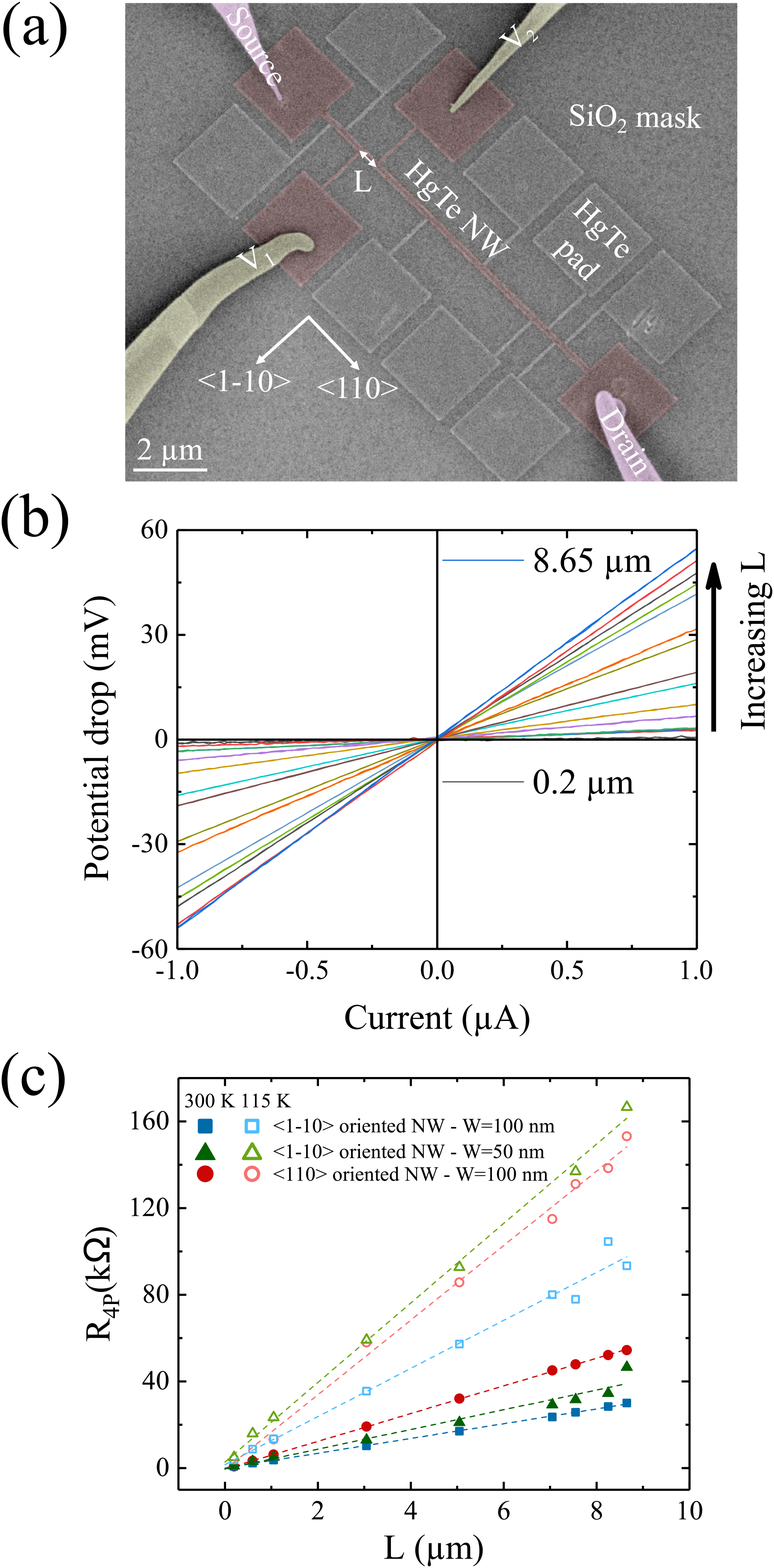}
   \caption{(a) SEM image of a $<$110$>$ oriented HgTe NW connected to HgTe pads grown by selective area molecular beam epitaxy on CdTe. The four STM tips and the HgTe nanostructures in contacts with the STM tips have been colorized. The tips used as potential probes are labelled V$_1$ and V$_2$. (b) Room temperature \textit{V(I)} characteristics measured for 14 different distances on the NW shown in (a). (c) Four-point resistance as a function of the NW length and two temperatures for three different NWs, the one oriented along $<$110$>$ corresponding to the one shown in (a). Dashed lines are linear fit curves.}
    \label{fig:FourProbes}
\end{figure}

\autoref{fig:Imagethick} displays the morphology evolution of $<$1$\bar{1}$0$>$ oriented NWs as function of the deposited nominal thickness. AFM cross sections taken from 50 nm- (top) and 100 nm-wide (bottom) openings at different nominal thicknesses are shown. The top side of the NW shrinks in size as the structure grows. Consequently the super-elevations might start to merge. It is the case for the 30 nm-thick growth of the 50 nm-wide NW (bright green top) and the 50 nm-thick growth of the 100 nm-wide NW (dark green bottom). Eventually, they would fully merge as shown by the 50 nm-thick growth of the 50 nm-wide NW (dark green top). Here, the cross section is almost a triangle with \{111\}A facets and a rounded tip. Due to the merging, we now have a height in the center exceeding the nominal thickness as described in \cite{Paper33}. The same trend is expected for the 100 nm wide NW with a greater deposited thickness. Considering a 10 nm deposition (light green), that is just above the mask level, only \{115\} facets are formed and lateral overgrowth is already present. We suppose that during the growth, the angle of the nanostructure facets will be more pronounced until it reaches $54.7$\textdegree, the angle of \{111\} facets. The width at the basis of the NW is constant with the deposited thickness. Nonetheless, after the merging, the lateral overgrowth might be larger as explained in \cite{Paper33}, but more investigation is needed to confirm this phenomenon.

To further assess the structural quality of the NWs, transport measurements were performed on NWs grown at GT = 180\textdegree C. As discussed above, this temperature limits the accumulation of materials on the edge of the NWs, providing a NW section which is close to the nominal surface, as defined in \autoref{fig:Imagerho}(a). 12 µm-long NWs with a nominal thickness of 30 nm were considered and connected to 12 lateral pads. The CdTe substrate was semi-insulating ensuring a conduction through the NW only. \autoref{fig:FourProbes}(a) shows a SEM image of one of the tip arrangements used to measure the resistance as a function of the distance between the NWs. The outer tips are in contact with the source and drain pads, in line with the NW, whereas the potential tips, V$_1$ and V$_2$, are in contact with opposite lateral pads, measuring the potential drop for a distance over NW segments of various lengths. Whatever the connected lateral pads, the \textit{V(I)} characteristics are linear, see \autoref{fig:FourProbes}(b), their slope yielding the four-point resistance, R$_{\mathrm{4p}}$. As seen in \autoref{fig:FourProbes}(c), R$_{\mathrm{4p}}$ is found to linearly increase with the distance at 300 K. This behavior is reproducible between the three NWs which have been investigated and indicates an homogeneous medium for the propagation of the charge carriers along the NWs. It is confirmed by the transport measurements performed at 115 K, although some data points are less aligned with the fitted curves because of unstable electric contacts between the cold sample and the STM tips maintained at room temperature. 
The linear increase of R$_{\mathrm{4p}}$ with the length \textit{L} of the NW is consistent with a diffusive transport, where the conductivity of the NWs is given by $\sigma  = (L/\mathrm{R_{4p}})/(t\cdot \mathrm{W})$. \autoref{tab:conduc} summarizes the conductivity measured for the three NWs in \autoref{fig:FourProbes}(b). The cross-section of NWs in both directions is deduced from the height profiles measured with the AFM scans as in \autoref{fig:Imagerho}(a). The conductivity shows a significant variation between both orientations, suggesting a lower structural disorder along the $<$1$\bar{1}$0$>$ direction. Such a result is consistent with the cross-sectional analysis of the $<$110$>$ oriented NWs, which has revealed the formation of side grains separated from the central grain by twin-boundaries. These twin-boundaries are known to act as scatterer and reduce the mobility of the charge carriers \cite{transporttwin1, transporttwin2}. Nevertheless, the experimental conductivity values at 300 K are significantly larger than that found in polycrystalline HgTe NWs grown from an Au nucleation site and only one order of magnitude lower than that of an MBE-grown HgTe film \cite{resisnorway}.
Further analysis of the conductivity as a function of the temperature reveals that the conductivity is smaller at 115 K. A thermally activated conductivity is the signature of a bulk energy gap in HgTe. Although the STEM characterization of the NW shown in \autoref{fig:ImageTEM1} and \autoref{fig:ImageTEM2} was not able to provide the precise strain undergone by HgTe, the transport measurements indicates the opening of a gap. This finding suggests that the HgTe NWs are strained on the CdTe substrate, in agreement with the in-plane strain occurring for HgTe films grown on CdTe \cite{bandgap1,bandgap2}.

\Table{\label{tab:conduc}Conductivity deduced from the length dependence of the four-point resistance for the three NWs measured in \autoref{fig:FourProbes}. W represents the opening width.}
\br
& & & & \multicolumn{2}{c}{R$_{\mathrm{4p}}$/\textit{L} (k$\Omega$/$\mu$m)} & \multicolumn{2}{c}{$\sigma$ (S/cm)} \\ \ms \cline{5-8} \ms
NW & Orientation & W (nm) & Area (nm$^2$) & 300 K & 115 K & 300 K & 115 K \\
\mr
1 & $<$1$\bar{1}$0$>$ & 100 & 4740 & $3.41\pm0.02$ & $11.02\pm0.55$ & $619\pm3$  & $191\pm9$ \\
2 & $<$1$\bar{1}$0$>$ & 50 & 2950 & $4.11\pm0.02$ & $18.35\pm0.41$  & $824\pm3$ & $184\pm4$ \\
3 & $<$110$>$ & 100 & 6260 & $6.40\pm0.02$ & $16.81\pm0.45$  & $249\pm1$ & $\095\pm3$ \\
\br
\endTable

\section{Conclusion}
In conclusion, we have demonstrated the SAG of in-plane HgTe nanostructures by MBE. The selectivity is achieved for several GTs and GRs with almost no parasitic growth on the mask, yielding a good reproduction of the designed patterns. A diffusion flux of Te adatoms from the mask area towards the openings, limited to less than 100 nm, influences the nanostructures growth along their edges and can be tuned by adjusting the GT. There is a strong anisotropy of growth morphology with respect to the three in-plane directions: $<$100$>$, $<$110$>$ and $<$1$\bar{1}$0$>$. While a flat top surface of NW is maintained for the $<$100$>$ NW direction, grains appear for the $<$110$>$ direction and super elevations for the $<$1$\bar{1}$0$>$ direction. For narrow wires, these super elevations merge into a peak with \{111\}A sides facets, yielding a triangular cross section with a rounded top tip. The thickness deposition can also be used to tune the morphology of the NWs. Above a certain thickness, at fixed width, the super-elevations can merge and the cross section turns again into a triangular shape with \{111\} sides facets. Finally, zincblende HgTe NW exhibits an epitaxial relation to CdTe(001) substrate. The NWs are found free of misfit dislocation despite the strain in the HgTe NW, which is evidenced by a thermally activated transport. 

\ack
The authors would like to acknowledge financial support
from the national research agency under the INSPIRING
project (ANR-21-CE09-0026-01). We also thank IEMN for the preparation of the samples.
Part of this work, carried out on the Platform for Nanocharacterisation (PFNC), was supported by the “Recherche Technologique de Base” program of the French National Research Agency (ANR). We also thank IEMN PCMP-PCP and CMNF platforms for the preparation and electrical characterization of the samples. J. X. acknowledges the financial support of the Fulbright-Tocqueville chair.

\section*{References}
\bibliography{document.bib}

\providecommand{\newblock}{}
\begin{thebibliography}{10}
\expandafter\ifx\csname url\endcsname\relax
  \def\url#1{{\tt #1}}\fi
\expandafter\ifx\csname urlprefix\endcsname\relax\def\urlprefix{URL }\fi
\providecommand{\eprint}[2][]{\url{#2}}

\bibitem{Paper1:1}
Kitaev A~Y 2003 {\em Annals of physics\/} {\bf 303} 2--30

\bibitem{Paper1:3}
Sarma S~D, Freedman M and Nayak C 2015 {\em npj Quantum Information\/} {\bf 1} 1--13

\bibitem{MolenkampScience}
Konig M, Wiedmann S, Brune C, Roth A, Buhmann H, Molenkamp L~W, Qi X~L and Zhang S~C 2007 {\em Science\/} {\bf 318} 766--770

\bibitem{candice1}
Ballet P, Thomas C, Baudry X, Bouvier C, Crauste O, Meunier T, Badano G, Veillerot M, Barnes J, Jouneau P~H {\em et~al.\/} 2014 {\em Journal of electronic materials\/} {\bf 43} 2955--2962

\bibitem{haas}
Haas B, Thomas C, Jouneau P~H, Bernier N, Meunier T, Ballet P and Rouvi{\`e}re J~L 2017 {\em Applied Physics Letters\/} {\bf 110}

\bibitem{Paper17:33}
Sarma S~D and Pan H 2021 {\em Physical Review B\/} {\bf 103} 195158

\bibitem{Paper17:34}
Ahn S, Pan H, Woods B, Stanescu T~D and Sarma S~D 2021 {\em Physical Review Materials\/} {\bf 5} 124602

\bibitem{SpintroniqueLeti1}
Fu Y, Li J, Papin J, Noel P, Teresi S, Cosset-Ch{\'e}neau M, Grezes C, Guillet T, Thomas C, Niquet Y~M {\em et~al.\/} 2022 {\em Nano Letters\/} {\bf 22} 7867--7873

\bibitem{SpintroniqueLeti2}
Noel P, Thomas C, Fu Y, Vila L, Haas B, Jouneau P~H, Gambarelli S, Meunier T, Ballet P and Attan{\'e} J 2018 {\em Physical Review Letters\/} {\bf 120} 167201

\bibitem{HgTe}
Thomas C, Crauste O, Haas B, Jouneau P~H, B{\"a}uerle C, L{\'e}vy L, Orignac E, Carpentier D, Ballet P and Meunier T 2017 {\em Physical Review B\/} {\bf 96} 245420

\bibitem{bocquillonEBL}
Bendias K, Shamim S, Herrmann O, Budewitz A, Shekhar P, Leubner P, Kleinlein J, Bocquillon E, Buhmann H and Molenkamp L~W 2018 {\em Nano letters\/} {\bf 18} 4831--4836

\bibitem{JosephsonHgTe}
Himmler W, Fischer R, Barth M, Fuchs J, Kozlov D~A, Mikhailov N~N, Dvoretsky S~A, Strunk C, Gorini C, Richter K {\em et~al.\/} 2023 {\em Physical Review Research\/} {\bf 5} 043021

\bibitem{HgTeEBL}
Ziegler J, Kozlovsky R, Gorini C, Liu M~H, Weish{\"a}upl S, Maier H, Fischer R, Kozlov D~A, Kvon Z~D, Mikhailov N {\em et~al.\/} 2018 {\em Physical Review B\/} {\bf 97} 035157

\bibitem{fu}
Fu L and Kane C~L 2008 {\em Physical review letters\/} {\bf 100} 096407

\bibitem{kitaev}
Kitaev A~Y 2001 {\em Physics-uspekhi\/} {\bf 44} 131

\bibitem{KrogstrupNature}
Krogstrup P, Ziino N, Chang W, Albrecht S, Madsen M, Johnson E, Nyg{\aa}rd J, Marcus C~M and Jespersen T 2015 {\em Nature materials\/} {\bf 14} 400--406

\bibitem{Paper2:5}
Deng M, Yu C, Huang G, Larsson M, Caroff P and Xu H 2012 {\em Nano letters\/} {\bf 12} 6414--6419

\bibitem{Paper2:6}
Churchill H, Fatemi V, Grove-Rasmussen K, Deng M, Caroff P, Xu H and Marcus C~M 2013 {\em Physical Review B\/} {\bf 87} 241401

\bibitem{norway}
Selvig E, Hadzialic S, Skauli T, Steen H, Hansen V, Trosdahl-Iversen L, van Rheenen A, Lorentzen T and Haakenaasen R 2006 {\em Physica Scripta\/} {\bf 2006} 115

\bibitem{norway2}
Haakenaasen R, Selvig E, Hadzialic S, Skauli T, Hansen V, Tibballs J, Trosdahl-Iversen L, Steen H, Foss S, Taft{\o} J {\em et~al.\/} 2008 {\em Journal of electronic materials\/} {\bf 37} 1311--1317

\bibitem{Paper6:13}
Kang J~H, Cohen Y, Ronen Y, Heiblum M, Buczko R, Kacman P, Popovitz-Biro R and Shtrikman H 2013 {\em Nano letters\/} {\bf 13} 5190--5196

\bibitem{Paper6:14}
Car D, Wang J, Verheijen M~A, Bakkers E~P and Plissard S~R 2014 {\em Advanced Materials\/} {\bf 26} 4875--4879

\bibitem{Paper1}
Aseev P, Fursina A, Boekhout F, Krizek F, Sestoft J~E, Borsoi F, Heedt S, Wang G, Binci L, Mart{\'\i}-S{\'a}nchez S {\em et~al.\/} 2018 {\em Nano letters\/} {\bf 19} 218--227

\bibitem{Paper2}
Lee J~S, Choi S, Pendharkar M, Pennachio D~J, Markman B, Seas M, Koelling S, Verheijen M~A, Casparis L, Petersson K~D {\em et~al.\/} 2019 {\em Physical Review Materials\/} {\bf 3} 084606

\bibitem{Paper5}
Desplanque L, Bucamp A, Troadec D, Patriarche G and Wallart X 2019 {\em Journal of Crystal Growth\/} {\bf 512} 6--10

\bibitem{Paper6}
Krizek F, Sestoft J~E, Aseev P, Marti-Sanchez S, Vaitiek{\.e}nas S, Casparis L, Khan S~A, Liu Y, Stankevi{\v{c}} T, Whiticar A~M {\em et~al.\/} 2018 {\em Physical review materials\/} {\bf 2} 093401

\bibitem{Paper9}
Desplanque L, Bucamp A, Troadec D, Patriarche G and Wallart X 2018 {\em Nanotechnology\/} {\bf 29} 305705

\bibitem{Paper15}
Jiang Y, Yang S, Li L, Song W, Miao W, Tong B, Geng Z, Gao Y, Li R, Chen F {\em et~al.\/} 2022 {\em Physical Review Materials\/} {\bf 6} 034205

\bibitem{Paper17}
Jung J, Schellingerhout S~G, Ritter M~F, ten Kate S~C, van~der Molen O~A, de~Loijer S, Verheijen M~A, Riel H, Nichele F and Bakkers E~P 2022 {\em Advanced Functional Materials\/} {\bf 32} 2208974

\bibitem{MarcusPRL}
Vaitiek{\.e}nas S, Whiticar A~M, Deng M~T, Krizek F, Sestoft J~E, Palmstr{\o}m C, Mart{\'\i}-S{\'a}nchez S, Arbiol J, Krogstrup P, Casparis L {\em et~al.\/} 2018 {\em Physical review letters\/} {\bf 121} 147701

\bibitem{Paper32}
Cachaza M~E, Christensen A~W, Beznasyuk D, S{\ae}rkj{\ae}r T, Madsen M~H, Tanta R, Nagda G, Schuwalow S and Krogstrup P 2021 {\em Physical Review Materials\/} {\bf 5} 094601

\bibitem{Paper33}
Morgan N, Dubrovskii V~G, Stief A~K, Dede D, Sangl{\'e}-Ferri{\`e}re M, Rudra A, Piazza V and Fontcuberta~i Morral A 2023 {\em Crystal Growth \& Design\/}

\bibitem{Paper36}
Dede D, Glas F, Piazza V, Morgan N, Friedl M, G{\"u}niat L, Dayi E~N, Balgarkashi A, Dubrovskii V~G and i~Morral A~F 2022 {\em Nanotechnology\/} {\bf 33} 485604

\bibitem{Paper40}
Dubrovskii V~G 2023 {\em Physical Review Materials\/} {\bf 7} 026001

\bibitem{infrared}
Ballet P, Baudry X, Polge B, Brellier D, Merlin J and Gergaud P 2013 {\em Journal of electronic materials\/} {\bf 42} 3133--3137

\bibitem{Paper5:12}
Desplanque L, Fahed M, Han X, Chinni V, Troadec D, Chauvat M, Ruterana P and Wallart X 2014 {\em Nanotechnology\/} {\bf 25} 465302

\bibitem{Paper29}
Iwanaga H, Tomizuka A, Shibata N and Mochizuki K 1986 {\em Journal of crystal growth\/} {\bf 74} 113--117

\bibitem{Paper38}
Albani M, Bergamaschini R, Salvalaglio M, Voigt A, Miglio L and Montalenti F 2019 {\em physica status solidi (b)\/} {\bf 256} 1800518

\bibitem{Paper20}
Oehling S, Ehinger M, Gerhard T, Becker C, Landwehr G, Schneider M, Eich D, Neureiter H, Fink R, Sokolowski M {\em et~al.\/} 1998 {\em Applied physics letters\/} {\bf 73} 3205--3207

\bibitem{Paper21}
Oehling S, Ehinger M, Spahn W, Waag A, Becker C and Landwehr G 1996 {\em Journal of applied physics\/} {\bf 79} 748--751

\bibitem{moire1}
Wang Y, Bruley J, Van~Meer H, Li J, Domenicucci A, Murray C~E and Rouviere J 2013 {\em Applied Physics Letters\/} {\bf 103}

\bibitem{moire2}
Beznasyuk D, Stepanov P, Rouvi{\`e}re J~L, Glas F, Verheijen M, Claudon J and Hocevar M 2020 {\em Physical Review Materials\/} {\bf 4} 074607

\bibitem{Paper41}
De~Donno M, Albani M, Bergamaschini R and Montalenti F 2022 {\em Physical Review Materials\/} {\bf 6} 023401

\bibitem{transporttwin1}
Kole{\'s}nik-Gray M~M, Hansel S, Boese M and Krsti{\'c} V 2015 {\em Solid State Communications\/} {\bf 202} 48--51

\bibitem{transporttwin2}
Qian X, Kawai M, Goto H and Li J 2015 {\em Computational Materials Science\/} {\bf 108} 258--263

\bibitem{resisnorway}
Gundersen P, Kongshaug K~O, Selvig E and Haakenaasen R 2010 {\em Journal of Applied Physics\/} {\bf 108}

\bibitem{bandgap1}
Br{\"u}ne C, Liu C, Novik E, Hankiewicz E, Buhmann H, Chen Y, Qi X, Shen Z, Zhang S and Molenkamp L 2011 {\em Physical Review Letters\/} {\bf 106} 126803

\bibitem{bandgap2}
Wu S~C, Yan B and Felser C 2014 {\em Europhysics Letters\/} {\bf 107} 57006

\end{thebibliography}

\end{document}